\documentclass[aps,amsmath, amssymb, english, aps, prl, twocolumn,reprint,floatfix, superscriptaddress,longbibliography]{revtex4-2}

\usepackage[english]{babel}
\usepackage{graphicx}
\usepackage{verbatim}

\usepackage{epsfig}
\usepackage{epstopdf}
\usepackage{amsfonts}
\usepackage{amsthm}
\usepackage{amsmath}
\usepackage{amssymb}
\usepackage{color}
\usepackage[usenames,dvipsnames,svgnames,table]{xcolor}
\usepackage{hyperref} 
\hypersetup{backref,pdfpagemode=FullScreen,colorlinks=true,linkcolor=NavyBlue,citecolor=BrickRed}
\usepackage{makeidx}
\usepackage{pifont}
\usepackage[normalem]{ulem}
\usepackage{soul}
\usepackage{color}

\usepackage{color}
\usepackage{grffile}


\def\expect#1{\mathinner{\langle{#1}\rangle}}

{\catcode`\|=\active
  \gdef\expect#1{\left<\mathcode`\|"8000\let|\bravert {#1}\right>}}
\def\bravert{\egroup\,\vrule\,\bgroup}

\def\beq{\begin{equation}}
\def\eeq{\end{equation}}
\def\be{\begin{equation}}
\def\ee{\end{equation}}

\def\cG0{{\cal G}_0}

\def\a{\alpha}

\def\uc2{$U_{c2}$}
\def\uc1{$U_{c1}$}

\def\bea{\begin{eqnarray}}
\def\eea{\end{eqnarray}}
\def \bal{\begin{align}}
\def \eal{\end{align}} 
\def\#{\!\!}
\def\@{\!\!\!\!}

\def\+{\dagger}

\begin{document}

\title{\bf Paradigm for finding d-electron heavy fermions: the case of Cr-doped CsFe$_2$As$_2$}

\author{Matteo Crispino}
\affiliation{Laboratoire de Physique et Etude des Mat\'eriaux, UMR8213 CNRS/ESPCI/UPMC, Paris, France}
\affiliation{Institut für Theoretische Physik und Astrophysik and Würzburg-Dresden Cluster of Excellence ct.qmat, Universität Würzburg, 97074 Würzburg, Germany}

\author{Pablo Villar Arribi}
\affiliation{Laboratoire de Physique et Etude des Mat\'eriaux, UMR8213 CNRS/ESPCI/UPMC, Paris, France}
\affiliation{International School for Advanced Studies (SISSA), Via Bonomea 265, I-34136 Trieste, Italy}

\author{Anmol Shukla}
\author{Fr\'ed\'eric Hardy}
\author{Amir-Abbas Haghighirad}
\author{Thomas Wolf}
\author{Rolf Heid}
\author{Christoph Meingast}
\affiliation{Institute for Quantum Materials and Technologies (IQMT), Karlsruhe Institute of Technology, 76131 Karlsruhe, Germany}

\author{Tommaso Gorni}
\altaffiliation{Present address: CINECA National Supercomputing Center, Casalecchio di Reno, I-40033 Bologna, Italy}
\affiliation{Laboratoire de Physique et Etude des Mat\'eriaux, UMR8213 CNRS/ESPCI/UPMC, Paris, France}

\author{Adolfo Avella}
\affiliation{Dipartimento di Fisica "E. R. Caianiello", Universit\`a degli Studi di Salerno, I-84084 Fisciano, Italy}
\affiliation{CNR-SPIN, UoS di Salerno, I-84084 Fisciano (SA), Italy}
\affiliation{Unit\`a CNISM di Salerno, Universit\`a degli Studi di Salerno, I-84084 Fisciano (SA), Italy}

\author{Luca de'~Medici}
\affiliation{Laboratoire de Physique et Etude des Mat\'eriaux, UMR8213 CNRS/ESPCI/UPMC, Paris, France}


\begin{abstract}
We define a general strategy for finding new heavy-fermionic materials without rare-earth elements: doping a Hund metal with pronounced orbital-selective correlations towards half-filling. We argue that in general band structures a possible orbital-selective Mott transition is frustrated by inter-orbital hopping into heavy-fermion behaviour - where d-orbitals provide both the heavy and the light electrons - which is enhanced when approaching half-filling. This phase ultimately disappears due to magnetic correlations, as in a standard Doniach diagram. Experimentally we have further hole doped CsFe$_2$As$_2$, a Hund metal with 0.5 electrons/Fe away from half-filling, and obtained a heavy fermionic state with the highest Sommerfeld coefficient for Fe-pnictides to date (270 mJ/mol K$^2$), before signatures of an antiferromagnetic phase set in.
\end{abstract}

\maketitle


Heavy-fermion materials have represented a major focus of the community working on materials with strong electronic correlations since their discovery in the seventies\cite{Andres_Heavy-fermions, Steglich_CaCu2Si2, Mott_HeavyFermions_SmB6, Doniach_his_diagram, Fulde-Heavy-fermions, Custers_Heavy-Fermions_QCP, Qimiao_Si_Steglich-HeavyFermion-Science_Review, Wirth_Steglich-Review_HeavyFermions}. 
Their specificity is a bad metallic state around room temperature harbouring free magnetic moments due to localized electrons, which disappear at cryogenic temperatures where a normal metal forms, a Fermi liquid. The latter, however, exhibits an electronic specific heat, which varies linearly at low temperatures in a standard metal, i.e. $C(T)\sim\gamma T$, where the "Sommerfeld coefficient" $\gamma$ is hugely enhanced compared to more conventional metals. As an example, in the heavy-fermion CeCu$_6$, $\gamma$ reaches 1600 mJ/mol~K$^{2}$, whereas for simple fcc Cu its value is $0.695$. $\gamma$ is indeed proportional to the effective mass of the electrons at the Fermi level, hence the label "heavy"\cite{Coleman_Book}.

Experimental and theoretical developments first obtained in the context of heavy-fermion research were instrumental for the enormous body of activity on unconventional, high-temperature superconductivity that followed the discovery of the cuprates at the end of the eighties, and that still propels the field today. As a matter of fact, the heavy fermion material CeCu$_2$Si$_2$ is the first discovered unconventional superconductor\cite{Steglich_CaCu2Si2}, i.e. one where the mechanism pairing the electrons into the Cooper pairs necessary for superconductivity is not based on phonons. Nowadays hundreds of heavy-fermions are known, and besides being a fertile ground for the exploration of the exotic physics of strongly correlated electrons, their intrinsic interest revolves around the interplay between electric and magnetic properties, and in the switching of properties that can be controlled with temperature, pressure or doping.

Key components for the overwhelming majority of these materials are f-electron elements, which naturally provide very narrow orbitals in which electrons easily localize. However, these elements, most of the time come with drawbacks: indeed Rare-Earths and Actinides are often rare (albeit not all as rare as their name implies), can be radioactive, and hard to extract or purify. Or simply, given the strategic value of many of them, they can be commercially problematic to obtain\cite{Van-Gosen-Rare_earths_critical}.

Nevertheless, serendipitous discoveries of heavy-fermion behaviour in d-electron compounds leave hope of exploring and exploiting heavy-fermion physics in f-element-free materials. The most remarkable case is that of LiV$_2$O$_4$\cite{Kondo_LiV2O4_heavyFermion} where $\gamma$=420~mJ/mol~K$^{2}$ approaches the Joule range of the materials with the heaviest f-electrons. The runner-ups,  YMn$_2$Zn$_{20}$\cite{Okamoto_YMn2Zn20_Heavy-Fermion} and Ca$_{1.5}$Sr$_{0.5}$RuO$_4$\cite{Nakatsuji_CaSr2RuO4_heavy}, are the only others with $\gamma > $200~mJ/mol~K$^{2}$, to the authors' best knowledge.  The mechanism behind this physics, (geometrical frustration\cite{Urano_LiV2O4_heavyFerm_frustration}, proximity to a Mott insulating state\cite{Arita_LiV2O4_hybridization}) is, however, still debated\cite{Miyazaki-Origin-d-electron-heavy-fermions}.

Here we outline a general strategy for the quest for new d-electron heavy fermions, and apply it to the very much studied class of Fe-based superconductors (FeSC).
Indeed the hole-overdoped stoichiometric end-members of the so-called "122" family, KFe$_2$As$_2$, RbFe$_2$As$_2$ and CsFe$_2$As$_2$, remarkably show enhanced Sommerfeld coefficient ($\sim$103, 127 and 180~mJ/mol~K$^{2}$, respectively) and a Curie-Weiss magnetic susceptibility (above a typically low coherence temperature), the hallmark of disordered local magnetic moments\cite{Hardy_KFe2As2_Heavy_Fermion, Hardy_122_SlaveSpin_exp, Eilers_Qcrit}. We here show how to enhance this behaviour, and that the paradigm on which the enhancement is based is easily generalized to a wide class of transition metal compounds.

For many of the main stoichiometric Fe-pnictide superconductors (with BaFe$_2$As$_2$ for the 122 family) the measured effective masses in the normal paramagnetic phase are actually only around 2-3 times the band mass\cite{Hardy_SpecHeat_Co122,Qazilbash_correlations_pnictides,Yi_Shen_ARPES_hole_electron_doped122}, which makes them moderately correlated metals.
However these mass enhancements differ for each band crossing the Fermi level, depending on their respective orbital character. This \emph{orbital selectivity} of the correlation strength\cite{demedici_3bandOSMT} is even more pronounced in the iron chalcogenides\cite{Yi_Universal_OSM_Chalcogenides,Kostin_FeSe_Orbital-selective_norm} which are more strongly correlated than iron pnictides.

This behaviour is perfectly captured by realistic theoretical calculations in which multi-orbital local dynamical correlations between conduction electrons are taken into account\cite{Haule_pnictides_NJP,YuSi_LDA-SlaveSpins_LaFeAsO,Lanata_FeSe_LDA+Gutz,Aichhorn_FeSe,Ferber_LiFeAs_LDA_DMFT,demedici_OSM_FeSC}. In particular, these are heavily influenced by the intra-atomic exchange coupling, known to be responsible for the famous "Hund's rules" in atomic physics. This earned these materials the label of "Hund metals"\cite{Yin_kinetic_frustration_allFeSC}. 
Indeed, Hund's coupling causes the high-spin local configurations to prevail in the quantum fluctuations of the metallic state, and, even if a Fermi-liquid state is formed at low temperatures, it typically shows a low coherence temperature and an enhanced effective mass, owing to the difficulty of establishing coherence among these very constrained local configurations. 

For the same reason a crucial factor for metallicity is the filling of the conduction bands\cite{demedici_MottHund,demedici_Janus}. In FeSC, these mainly arise from the five d-orbitals of the transition metal atom. 
In BaFe$_2$As$_2$, 6 electrons/Fe populate the five conduction bands and the Fe atom is nominally in a $d^6$ configuration, whereas in the aforementioned end-members AFe$_2$As$_2$ (A=K,Rb,Cs) the Ba cation is replaced by an alkali atom, which introduces half a hole per Fe/atom, and a $d^{5.5}$ configuration is reached.
By doping more holes, half-filling can be reached (in practice this needs the substitution of other atoms in the unit cell) and thus a $d^5$ configuration, where, mainly, each one of the five d-orbitals is occupied by one of the 5 electrons, so as to maximize the total spin. No orbitals are empty or doubly occupied, and thus the quenched orbital degree of freedom cannot fluctuate and contribute to delocalized quasiparticle excitations. Metallicity is thus minimal under these conditions, which is typically enough to obtain a Mott insulator in 3d compounds\cite{An_Sefat-BaMn2As2_ins,McNally_BaMn2As2_Mott_Hund}. 
\begin{figure}[h!]
\begin{center}
\includegraphics[width=0.48\textwidth]{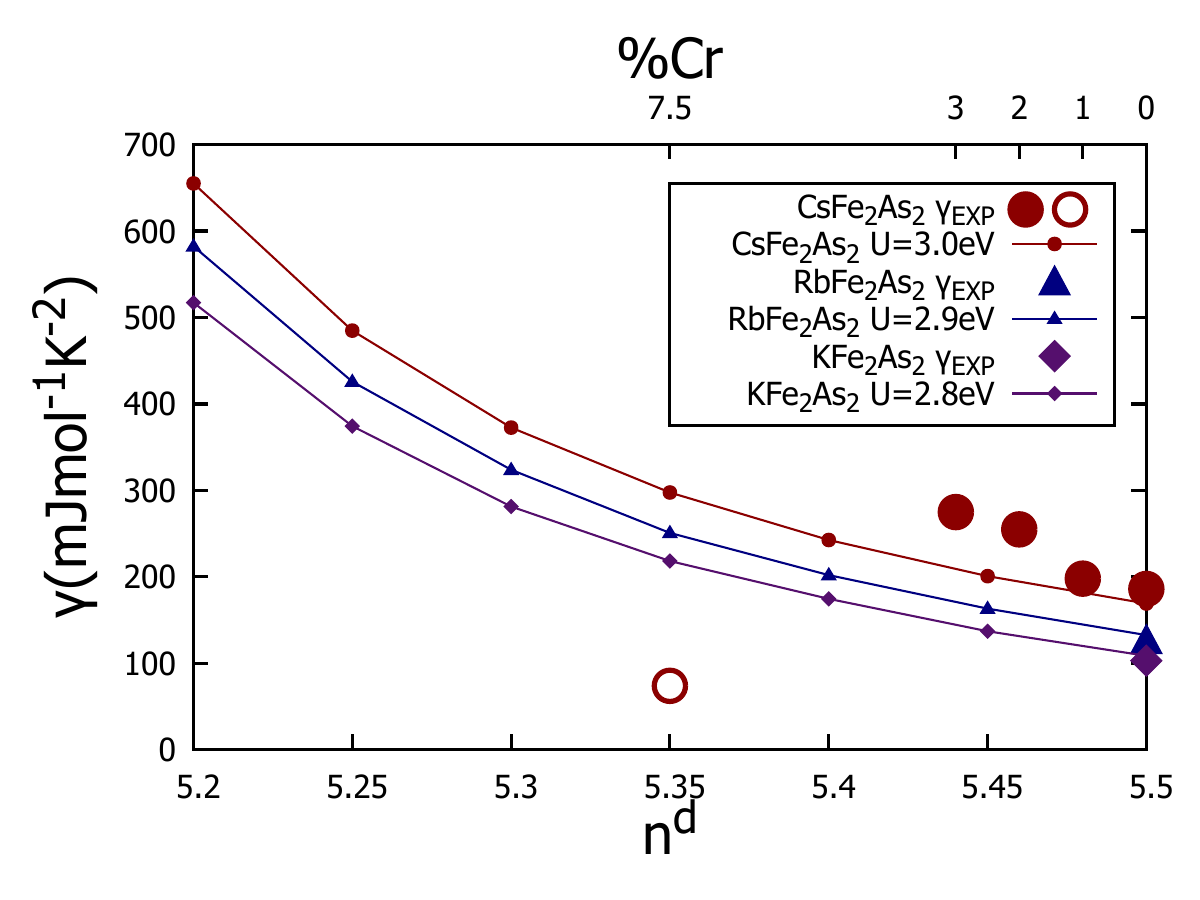}
\caption{Increase of the specific-heat Sommerfeld coefficient as a function of hole-doping in AFe${_2}$As${_2}$ (A=\{K, Rb, Cs\}), calculated within Density-Functional Theory+Slave-spin mean-field (lines, see Methods). Large symbols report the measured experimental values for Cr-doped CsFe${_2}$As${_2}$.}\label{fig:SSMF_Sommerfeld}
\end{center}
\end{figure}
Correlations are therefore expected to increase when moving towards half-filling\cite{demedici_Hunds_metals,Hardy_122_SlaveSpin_exp,Backes_KRbCs122}, and this is indeed observed in Ba$_{1-x}$K$_x$Fe$_2$As$_2$ in the whole range $0<x<1$ \cite{Hardy_122_SlaveSpin_exp}. 
Experiments and theory go hand-in-hand in showing that the orbital selectivity is also enhanced along with the correlation strength\cite{demedici_OSM_FeSC} within the same range. 
Further increase of corrrelations and selectivity can be obtained with chemical pressure. Indeed, the sequence of isovalent substitutions of K with Rb and Cs (having the main effect of bringing the Fe atoms further apart in the plane\cite{Eilers_Qcrit}) leads to the strongest mass enhancements (which are measured for the bands of orbital character $d_{xy}$) from a factor $\sim$~20 to $\sim$~40\cite{Hardy_122_SlaveSpin_exp}. 

Despite these enormous enhancements, the electronic masses do not actually diverge, which would imply what is known as an orbital-selective Mott transition (OSMT). After such a transition, which was the object of several model studies\cite{Anisimov_OSMT,Koga_OSMT,demedici_Slave-spins,demedici_3bandOSMT}, the heaviest electrons would be localized due to the strong local correlations, while the others would remain itinerant. 
However in the studied models it was shown that a hopping amplitude hybridizing the orbitals is enough to prevent this selective localization\cite{demedici_Cerium, koga05, demedici_Slave-spins,Kugler_Kotliar-OSMT_hyb_NRG}. Here we show that this is completely general: in a realistic band structure local correlations cannot induce an OSMT at zero temperature.
The argument is the following. Consider a Fermi liquid with a generic band structure in which electrons of one orbital character are considerably closer to Mott localization. This can be expressed by their quasiparticle weight Z (which is the inverse of the mass enhancement when correlations are local) being very small compared to the others. Then a vanishing of the Z would cause all the (intra and inter-orbital) hopping integrals of electrons with that orbital character vanish as well, producing both the flattening and the de-hybridization of a band, thus decoupling that band from the band structure. This hybridized structure entails, however, a singular contribution $\propto Z\log(Z)$  to the kinetic energy of the system (see Supplementary Material). In a variational description of this Fermi-liquid phase (such as the Gutzwiller approximation or an equivalent slave-particle mean field\cite{kotliar_ruckenstein,Crispino-Broken_Sym_SSMF-Neel}) it is obvious that the minimum of the total energy (which has also a regular interaction part) cannot be in Z=0, where the kinetic energy has an infinite slope as a function of Z. Thus the de-hybridization and the OSMT are prevented. 

\subsection{Frustrated orbital-selective Mott transition}

Such a \emph{frustrated OSMT}, however, can be precisely the source of the heavy-fermion behaviour we are looking for, in that it creates a situation where one part of the d-electrons is sensibly heavier than - but still hybridized to - the rest of the system, thus mimicking the role of the f-electrons in the original heavy fermion compounds. On the other hand the mechanism frustrating their complete localization guarantees a large region of parameters where Z is small but finite, a situation that would otherwise require fine tuning of the composition, or of external thermodynamic parameters such as pressure, magnetic field, etc.

In the hope of exploiting this resistance to localization, we have experimentally hole doped the iron pnictide with the largest $\gamma$, CsFe${_2}$As${_2}$, by partially substituting Fe with Cr atoms. Here, we take advantage that this substitution hardly changes the lattice parameters \cite{Sefat_Cr-dop_BaFe2As2,Li23_Cr_dopedCs122}, providing a thorough way of disentangling charge-doping and chemical-pressure effects. Thus, a further enhancement of the correlations is expected as Cs(Fe$_{1-x}$Cr$_{x}$)${_2}$As${_2}$ is approaching half-filling at $x$ = 0.25 (each Cr substitution providing 2 extra holes), leaving the rest of the electronic band structure virtually unaltered. As illustrated in Fig.\ref{fig:SSMF_Sommerfeld}, our calculations, assuming the persistence of a paramagnetic ground state, show a substantial increase of the Sommerfeld coefficient with just a few \% of Cr in AFe${_2}$As${_2}$ (A=\{K, Rb, Cs\}).

\subsection{Experimental evidence for heavy-fermion behavior in Cr-doped CsFe$_2$As$_2$}

\begin{figure*}[t]
\begin{center}
\includegraphics[width=\textwidth]{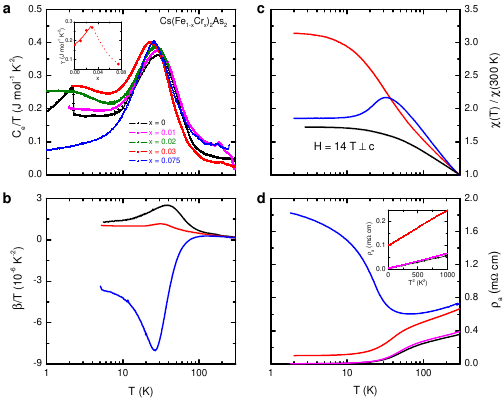}
\caption{\textbf{Thermodynamic and transport properties of Ba(Fe$_{1-x}$Cr$_x$)$_2$As$_2$}. (a) Temperature dependence of the electronic heat capacity C$_e$(T) = C(T) - C$_{lat}$(T) for various Cr substitution. Here, C$_{lat}$(T) represents the phonon contribution inferred from DFT calculations (see Fig.~S4a in the Supplementary Material). The inset shows the evolution of the Sommerfeld coefficient, $\gamma$ = $\left(\frac{C_e}{T}\right)_{T \rightarrow 0}$, with increasing $x$. Lines are guides to the eye. (b) Corresponding volume thermal-expansion coefficient, $\beta(T)$ = 2$\alpha_a$(T) + $\alpha_c$(T), as a function of temperature. The $a$- and $c$-axis uniaxial contributions $\alpha_a$(T) and $\alpha_c$(T) are shown in the Supplementary Material (Fig. S3a and S3b). (c) and (d) show the temperature dependence of the magnetic susceptibility $\chi$(T) measured for H $\perp$ $c$ (in a constant field of H = 14 T) and the zero-field $a$-axis resistivity $\rho_a$(T), respectively. The inset shows the T$^2$ dependence of $\rho_a$, indicative of a heavy Fermi-liquid behavior for $x \leq 0.03$.}\label{fig:Exp1}
\end{center}
\end{figure*}

Figure \ref{fig:Exp1} summarizes our thermodynamic and transport measurements performed on Cs(Fe$_{1-x}$Cr$_{x}$)${_2}$As${_2}$ single crystals. The electronic heat capacity $C_{e}$(T) (obtained by subtracting, from the measured data, the phonon contribution inferred from Density-Functional Theory (DFT) calculations - see Methods) together with the volume thermal expansion $\beta$(T) (which is the sum $\beta(T)=\sum_i \a_i(T)$ of the linear thermal-expansion coefficients $\a_i(T)=1/L_i(\frac{\partial L_i}{\partial T})_{p_i}$, along the three axes $i=x,y,z$) are illustrated in Figs. \ref{fig:Exp1}(a)-(b), respectively. Both C$_{e}$/T and $\beta$/T tend towards constant values at low temperature, as expected for a Fermi liquid. This is further supported by the Pauli-like susceptibility and the $AT^2$ dependence of the resistivity for $x$ $\leq$ 0.03, as shown in \ref{fig:Exp1}(c)-(d), respectively. The Sommerfeld coefficient $\gamma$ = $\left(C/T \right)_{T \rightarrow 0}$  is large and strongly increases from 180 mJ /(mol K${^2}$) for $x$ = 0 to about 270 mJ/(mol K${^2}$) at $x$ = 0.03 (which corresponds to 6 \% extra holes per transition-metal atom). Concomitantly $A$ is enhanced by a factor of $\approx$ 3 (see inset of Fig.\ref{fig:Exp1}(d)), thus maintaining a roughly constant Kadowaki-Woods ratio $A/\gamma^2$.  
We note that this large $\gamma$ is among the largest observed in $d$-electron compounds, and the highest for Fe-based superconductors. For comparison, $\gamma$ falls in the range of $U$-based heavy-fermion materials, {\it e.g.} URhGe, UGe$_2$ and UTe$_2$ \cite{Aoki_review,Willa_UTe2}. Additional feature related to heavy-fermion physics is the pronounced extrema in Figs \ref{fig:Exp1}(a)-(b) around $T^*$ = 25 K which was previously reported in both (Ba$_{1-x}$K$_x$)Fe$_2$As$_2$ and AFe${_2}$As${_2}$ (A=\{K, Rb, Cs\}) \cite{Hardy_KFe2As2_Heavy_Fermion} and assigned to the archetypal coherence-incoherence crossover below which heavy quasiparticles are formed. $T^*$ shifts towards lower temperature with increasing hole content, and as noted for AFe${_2}$As${_2}$ it scales approximately with $1/ \gamma$. Further, the electronic Grüneisen parameter $\Gamma_e=B_T \left( \frac{\partial\ln{T^*}}{ \partial p}\right)$ can be evaluated as
\begin{equation}\label{eq:Gruneisen}
\Gamma_e\approx-B_T \left( \frac{\partial\ln{\gamma}}{ \partial p}\right) \approx V{_m} B{_T}\left(\frac{\beta(T)}{C(T)}\right)_{T \rightarrow 0},
\end{equation}
(where $B_T$ and V$_m$ are the bulk modulus and the molar volume, respectively, and where we have used $\beta(T)=-{\frac{T}{V_m}(\frac{\partial S}{\partial p})}_T$, from a thermodynamic Maxwell relation, and the Fermi-liquid relation $C\sim\gamma T$), which typically measures the volume dependence of the electronic correlations, amounts to +20 for $x$ = 0, a moderate value, however, comparable in magnitude to that of $U$-based Kondo lattices \cite{deVisser_URhGe,Willa_UTe2}. Here, the positive sign of $\Gamma_e$ indicates that $\gamma$ decreases with increasing hydrostatic pressure, {\it i.e.} quite naturally pressure suppresses correlations by increasing the bandwidth.

\subsection{Entering antiferromagnetism}

Interestingly, drastic changes occur for higher Cr concentration. $\gamma$ drops dramatically by a factor of more than 3 for $x$ = 0.075, while the resistivity now exhibits a moderate insulating-like behavior, consistent with recent measurements \cite{Li23_Cr_dopedCs122}. In parallel, the thermal expansion changes sign, revealing that correlations now increase with hydrostatic pressure with an enhanced and negative Grüneisen parameter $\Gamma_e$ $\approx$ - 86. This sign-changing  shows that the entropy is going over a maximum with increasing Cr content, as expected theoretically, {\it e.g.} in the vicinity of a quantum critical point \cite{Garst05}. Although $C_{e}(T)$ and $\alpha(T)$ do not exhibit sharp discontinuities characteristic of a phase transition, we find a negative peak near $T^*$ which is sharper than observed for $x$ $<$ 0.075. The magnetization at $x$ = 0.075 in H = 14 T is reminiscent of that of an antiferromagnet with an ordering temperature around 25 - 30 K (for fields applied perpendicular to the easy direction). We interpret $T^*$ as a broadened antiferromagnetic transition which may sharpen up and turn into a genuine phase transition for larger Cr substitution (or for samples with reduced disorder).  We note that a splitting of the field-cooled (FC) and zero-field-cooled (ZFC) susceptibility curves is observed around 25 K in our low-field susceptibility data (Fig.~S2c in the Supplementary Material), which was ascribed in Ref.\cite{Li23_Cr_dopedCs122} to a spin-glass transition emerging from frustrated magnetic interactions. However, a bifurcation of the FC and ZFC curves, resulting from domain-wall motion, were previously also observed in BaFe$_2$As$_2$ \cite{He_Ba122_FeSe} at the Néel temperature. Importantly, our data reveal that this glassy behavior is progressively shifted to lower temperature with increasing field strength. For $H$ = 14 T, this splitting occurs below 10 K and the susceptibility curves are fully reversible beyond, confirming that the peak at $T^*$ is indeed a bulk thermodynamic feature not related to glassy physics.

This magnetic scenario is in line with the increase of the NMR spin-lattice relaxation time observed near $T^*$ \cite{Li23_Cr_dopedCs122}, which typically indicates the emergence of sizable antiferromagnetic correlations, as observed for other Fe-pnictides and chalcogenides. This is further supported by the fact that both $d\rho/dT$ and $d(T\chi)/dT$, shown in Supplementary Material Fig.~S4, exhibit a closely similar behavior to that of the specific heat and the thermal expansion around 25 K. This is indeed expected in the vicinity of a magnetic transition, as predicted by Fisher and Langer {\it et al.} in Refs \cite{Fisher1, Fisher2}. Thus, our experimental results strongly suggest that the system is evolving towards magnetic ordering, most likely antiferromagnetism. Indeed, the proximity to magnetism provides a natural explanation for the observed negative Grüneisen parameter, {\it i.e.} an increase of the correlations, since hydrostatic pressure typically suppresses antiferromagnetism in itinerant systems, as previously reported for the spin-density wave transition in both (Ba$_{1-x}$K$_x$)Fe$_2$As$_2$ and Ba(Fe$_{1-x}$Co$_{x}$)$_2$As$_2$ \cite{Meingast12, Hassinger}.

\subsection{Heavy-fermion vs magnetic behaviour: Doniach-like phase diagram}

\begin{figure}[htb]
\begin{center}
\includegraphics[width=0.48\textwidth]{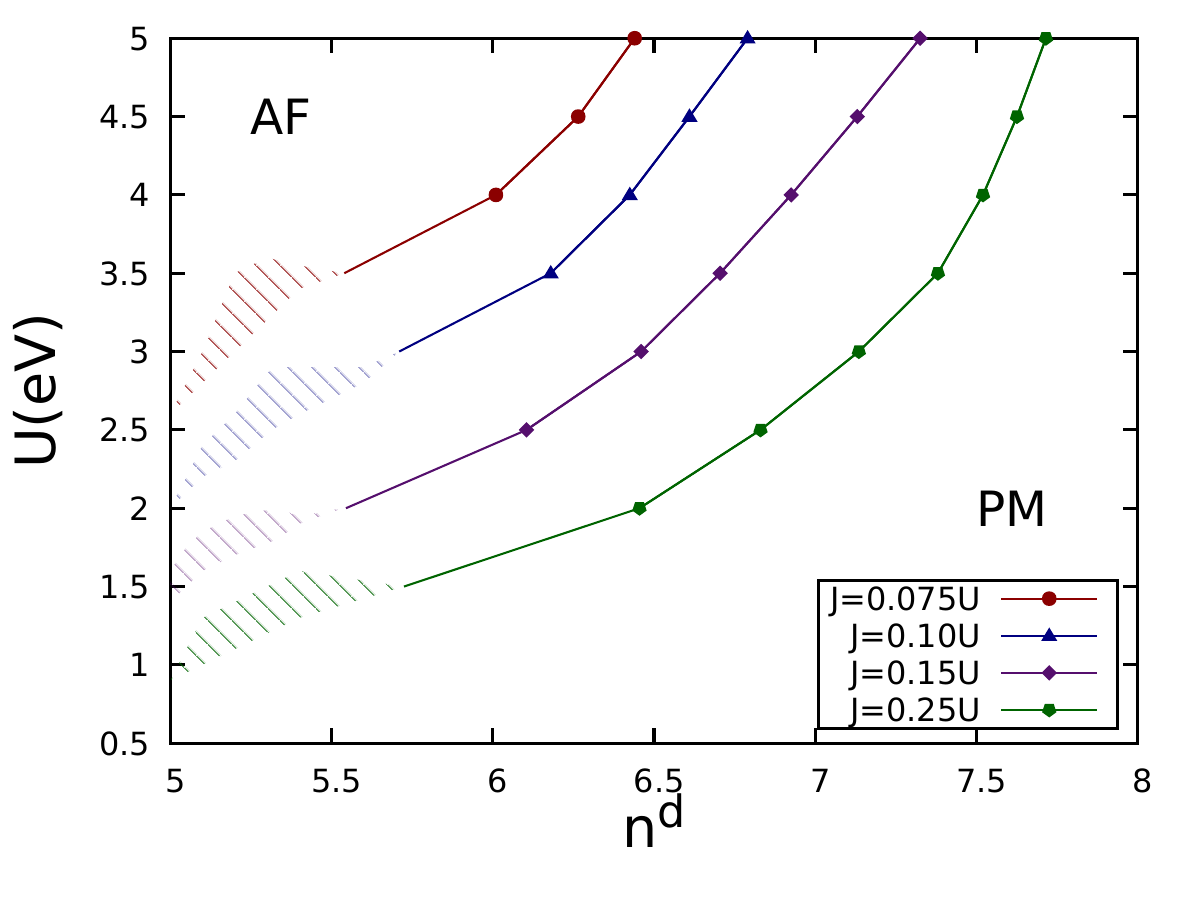}
\caption{Theoretical phase diagram of CsFe$_2$As$_2$ in the interaction-doping plane. Different values of $J/U$ are explored. In all cases a frontier divides a paramagnetic metallic phase (low U, large doping) from a G-type AF phase (high U, small doping). The transition can be 2nd-order (continuous line), or 1st-order (the hatched zones indicate roughly the associated zone of phase separation).}\label{fig:AF}
\end{center}
\end{figure}

Additional evidence for antiferromagnetism comes from our theoretical simulations. Indeed, towards half-filling, the paramagnetic state becomes unstable and the system is found to undergo a phase transition into a $G$-type antiferromagnetic (AF) state, as reported in Fig.~\ref{fig:AF}. This happens at a frontier, the exact position of which is obviously sensitive to the choice of the interaction parameters (the extent of the AF phase is most likely overestimated by the mean-field treatment we use here\cite{Misawa_d5-proximity_magnetic,Misawa_LaFeAsO}). However, in all cases the AF phase is accessed - and its staggered magnetization increases monotonically - by increasing the interaction strength or by doping towards half-filling (indeed a G-type antiferromagnetic phase was experimentally found near or at half filling in Refs.\cite{McNally_BaMn2As2_Mott_Hund,Singh_BaMn2As2,Marty_BaCr2As2}).
Now, the magnetic state reduces the quantum fluctuations, and correlations can be considerably reduced by an increasing magnetization. In realistic simulations, increasing the local interaction strength (parametrized by U in Fig.~\ref{fig:AF}) while keeping all other parameters fixed is a simple way to mimic a negative hydrostatic pressure (i.e. a reduction of all the hoppings), and indeed in the AF phase one can find zones where $\gamma$ decreases with increasing U. This implies a positive $\left( \frac{\partial\ln{\gamma}}{ \partial p}\right)$, and thus a negative thermal-expansion coefficient.
Theory thus further supports the measured $\Gamma_e <0$ as an evidence of the entrance into an AF phase, in accord with previous experimental analyses\cite{Kimber_NTE_AF_1111,Klimczuk_NTE_AF_1111,Hu_NTE_AF_CrAs}.

It is worth noting that our theory predicts this magnetic transition to be either 2nd-order or 1st-order (thus implying a zone of phase separation\cite{Crispino-Broken_Sym_SSMF-Neel}) depending finely on the interaction strength. An inhomogeneous phase could thus well be at the origin of the bad-metallic behaviour of the resistivity reported above. 

The heavy-fermion behaviour is thus enhanced by doping towards half-filling as predicted, but it is eventually cut off by the insurgence of a magnetic phase. This is generically in line with the typical "Doniach" diagram of f-electron heavy fermions\cite{Doniach_his_diagram}. However this analogy needs some context. For f-electron materials the Kondo coupling rules not only the Fermi-liquid coherence scale, but also the magnetic coupling, which is typically of the RKKY type. The two scales are bound to cross because of their different functional dependence on the Kondo coupling, which guarantees a transition to the magnetic states. In d-electron compounds magnetic couplings of different origins compete, and most likely superexchange dominates in the present case. Nevertheless this guarantees a transition as well, since with decreasing doping the Fermi-liquid scale is reduced, while the superexchange remains roughly constant, so that close enough to half-filling it will always dominate.

It is also worth discussing a variation of this scenario that has relevance for the standard heavy fermions. Indeed we have shown that hybridization of the heavy band with the rest of the system prevents local correlations to bring the system into an OSMT. However non-local correlations can succeed\cite{Paul_Pepin_Norman-Kondo_Breakdown,Pepin_OSMT,deLeo_PAM_CDMFT_PRL}, and cause Z to vanish before the mass diverges, since these two quantities are distinct in this case.
Local correlations being dominant in many d-electron materials, we expect that in many realistic situations a frustration of the OSMT will still persist enough to cause heavy-electron behaviour. If however non-local correlations manage to cause an OSMT, magnetism will immediately set in\cite{Biermann_nfl,deLeo_PAM_CDMFT_PRB,Bascones_OSMT_Gap_halffilling,Rincon_Dagotto-BlockMag_OSMT}, since the selective Mott state has free local moments whose excess entropy needs to be quenched. A phase diagram reproducing the so-called "local quantum critical point" scenario\cite{Si_Rabello_Ingersent_Local_QCP} is thus produced\cite{deLeo_PAM_CDMFT_PRB}, where both the Fermi-liquid scales and the magnetic-order vanish at the same (OSMT) point, in a typical fan-like fashion. Non-Fermi liquid behaviour is expected in the selective phase\cite{Biermann_nfl,Winograd_pseudogap}, which is then possibly observable in the "quantum critical" zone of the phase diagram.

In summary, we have outlined a general strategy for the search of new f-electron-free heavy-fermion materials: doping towards half-filling a Hund metal in order to enhance the orbital-selectivity of its electronic correlations. Before a magnetic phase sets in, extremely heavy mass enhancements can be reached for electrons with a given orbital character, while the others remain much less correlated, thus mimicking the f-electrons and the conduction electrons of traditional heavy-fermions, respectively.
We have applied our paradigm to the case of the hole-doped end-member of the 122 family of Fe-based superconductors CsFe$_2$As$_2$, that we have further hole-doped by substituting Fe with Cr. At 3\% concentration of Cr we have reached a Sommerfeld coefficient of $\gamma\simeq$~270~mJ/mol~K$^{2}$, a record value for FeSC to date.

\subsection{Methods}
Single crystals of Cs(Fe$_{1-x}$Cr$_x$)$_2$As$_2$ (with x= 0, 0.01, 0.02, 0.03,0.075)  were grown from  a Cs-rich self-flux in a sealed environment. Cs, Fe, Cr and As were weighted in molar ratio 8:1-x:x:11, respectively. All sample manipulations were performed in an argon glove box (O$_{2}$ content is $<$ 0.5 ppm). Molten Cs together with a mixture of FeAs$_{2}$ flux, chromium and arsenic were loaded into an alumina crucible. The alumina crucible with a lid was placed inside a stainless steel container and encapsulated. The stainless steel container was placed in a tube furnace filled with 300 mbar Argon gas and heated up to 200°C. The mixture was kept at this temperature for 10 h and subsequently heated up to 980°C -1050°C in 50°C/h. The furnace temperature was kept constant at 980°C-1050°C for 5 h and slowly cooled to 760 °C at the rate of 0.5°C/h to 3°C/h depending on the chromium content used for the growth and subsequently, the furnace was canted to separate the excess flux. After cooling to room temperature, shiny plate-like crystals with typical sizes 4 x 2 x 0.4 mm$^{3}$ were easily removed from the remaining ingot. Electron micro probe analysis was performed on Cs(Fe$_{1-x}$Cr$_x$)$_2$As$_2$ crystals using a compact scanning electron microscope (SEM) – energy dispersive x-ray spectroscopy (EDS) device COXEM EM-30$^{plus}$ equipped with an Oxford Peltier-cooled silicon drift detector. The EDS analyses on the Cs(Fe$_{1-x}$Cr$_x$)$_2$As$_2$ crystals revealed chromium content x = 0.0, 0.01, 0.03 and 0.075. A scanning electron micrograph and electron microanalysis pattern is shown in the Supplementary Material (Fig.~S5).  

To obtain an estimate for the phonon contribution to the specific heat, we performed first-principles calculations of the vibrational properties of Cs(Fe$_{1-x}$Cr$_x$)$_2$As$_2$ in the framework of the mixed-basis pseudopotential method \cite{Mixed_Basis}. Norm-conserving pseudopotentials were generated using the scheme of Vanderbilt \cite{Van_PRB}, and included semicore 3$s$, 3$p$-states for Fe and 5$p$-states for Cs.
For the mixed-basis expansion of the valence states, we combined plane-waves with a cutoff of 22Ry and local functions of $s$, $p$, and $d$-type at Fe sites and of $p$ and $d$-type at Cs sites. The PBE form of the generalized gradient approximation (GGA) was used for the exchange-correlation functional \cite{Perdew_PBE}. Brillouin zone summations were done with a Gaussian broadening technique using a broadening of 0.1 eV and 40 wave vector points in the irreducible part of the Brillouin zone. Using experimental lattice constants, vibrational properties were calculated via density function perturbation theory, as implemented in the mixed-basis method \cite{Heid_PRB}. Dynamical matrices were obtained on a 4x4x2 tetragonal momentum mesh, and standard Fourier interpolation techniques were applied to subsequently calculate the phonon density of states, from which the phonon specific heat was derived. The same phonon spectrum was used in previous publications \cite{Eilers_PRL,Wiecki_PRL}.

Thermal expansion was measured using a home-built high-resolution capacitance dilatometer \cite{Meingast_Dilatometry}. Heat-capacity, resistivity and magnetization measurements were carried out in Physical Properties Measurement System (PPMS) from Quantum Design.

The numerical calculations were performed using a combination of DFT and Slave-Spin Mean-Field method (SSMF), as implemented in Ref.\onlinecite{Crispino-Broken_Sym_SSMF-Neel}. The \emph{ab-initio} DFT simulation uses the PBE GGA exchange-correlation functional and the resulting conduction bands are expressed as a tight-binding model on a basis of local maximally-localized Wannier Functions centered on the Fe atoms. Dynamical electronic correlations result from the inclusion of standard multi-orbital Hubbard local-interaction Hamiltonian, in the density-density form of the Kanamori interaction (i.e. where pair-hopping and spin-flip terms are omitted). The SSMF is solved at zero temperature. In Fig.\ref{fig:SSMF_Sommerfeld} we fix $J=0.25U$ for all the studied compounds.  

\acknowledgements 
MC, PVA and LdM acknowledge A. Amaricci for discussions on the numerical implementations of the SSMF method.
MC, PVA, TG and LdM are supported by the European Commission through the ERC-CoG2016, StrongCoPhy4Energy, GA No724177. MC is supported by the W\"urzburg-Dresden Cluster of Excellence on Complexity and Topology in Quantum Matter –ct.qmat Project-ID 390858490-EXC 2147. AAv acknowledges support by MIUR under Project No. PRIN 2017RKWTMY. Work at KIT was partially funded by the Deutsche Forschungsgemeinschaft (DFG, German Research Foundation) TRR 288-422213477 (Project A02). A.S. acknowledges funding from the European Union’s Horizon 2020 research and innovation program under the Marie Skldowska-Curie Grant Agreement No. 847471 (QUSTEC).


\bibliography{bibldm,FeSC,publdm}

\end{document}


\title{Supplementary material for Paradigm for finding d-electron heavy fermions: the case of Cr-doped $\mathbf{CsFe_{2}As_{2}}$}


\author{Matteo Crispino}

\author{Pablo Villar Arribi}

\author{Anmol Shukla}
\author{Fr\'ed\'eric Hardy}
\author{Amir-Abbas Haghighirad}
\author{Thomas Wolf}
\author{Rolf Heid}
\author{Christoph Meingast}

\author{Tommaso Gorni}

\author{Adolfo Avella}

\author{Luca de'~Medici}


\maketitle

\setcounter{page}{\pagenumbaa}
\thispagestyle{plain}

\section{Frustrated orbital selective Mott transition}
We here demonstrate that, in a generic realistic bandstructure, local correlations cannot induce an orbital-selective Mott transition (OSMT) at zero temperature.

We start our reasoning by assuming the system to be in a Fermi liquid state, with a generic band structure (i.e. typically including - in a tight-binding representation - mostly nonzero hopping amplitudes, and in particular inter-orbital ones). We consider the situation in which the self-energy is diagonal in orbital space and the electrons of one particular orbital character are considerably closer to a Mott localisation than those of the other bands. That is, the quasiparticle weight $Z$ associated to this orbital character is very small compared to the others. If this quasiparticle weights could reach zero, it would result in an OSMT. If correlations are purely local (i.e. the self-energy is local and the quasiparticle weight coincides with the inverse mass enhancement) this implies the vanishing of both the intra-orbital and inter-orbital hoppings of the electrons with that orbital character. This leads to the flattening and the de-hybridisation of the corresponding band from the rest of the band structure. However, we are now going to show that the hybridised band structure entails a singular contribution in the kinetic energy of the system which vanishes proportionally to $Z\ln Z$, as $Z\rightarrow 0$. Then, in a variational description of the Fermi liquid (such as Gutzwiller approximation, slave-boson mean field, slave-spin mean field) this logarithmic singularity prevents the system's energy\footnote{The system's energy also contains an interaction term that we reasonably suppose to be regular.} to realize a minimum in $Z=0$ where the kinetic energy slope is infinite, and frustrates the de-hybridisation and the realization of an orbital selective Mott phase. We refer to this impossibility of OSMT to occur as \emph{frustrated orbital-selective Mott transition}(fOSMT)\footnote{We generalize here the model result of Ref.\cite{deMedici_Mott_Kondo}. Evidences of this behaviour can be found in Refs.\cite{demedici_Slave-spins,Koga_PRL_OSMT,Kugler_FOSMT}}.

Let us then consider a system of $M$ bands of mixed orbital character at \emph{zero temperature}. In k-space they can be written as a non-diagonal matrix in the orbital basis where the k-independent self-energy is diagonal. Let's explicitly signal only the renormalization factor Z of the most correlated orbital. We also set the latter to be half filled, which is ultimately a necessary condition for developing an OSMT. This implies that, in the $Z\rightarrow 0$ limit, the band mainly of this orbital character becomes flat and coincides with the Fermi level. Considering at first precisely the limiting situation $Z=0$, we are then left with a subset of $M-1$ bands that can be diagonalised, in order to obtain the $Z=0$ band structure. This $Z=0$ band structure has a Fermi surface, which is also obviously the \emph{locus} $\mathcal{K}_0$ of intersection of all the bands with the flat band in this limit. We denote the density of states at the Fermi level of this band structure (minus the flat band) as $D^{Z=0}(E_F)$.

For finite Z, in the new basis, the hopping matrix in k-space reads:
\begin{equation}
\label{eqn:Band_HopMatrix}
\left(
\begin{array}{ccc|c}
\epsilon_1(\mathbf{k}) & 0 & \dots & \sqrt{Z}V_{1M}(\mathbf{k}) \\
0 & \epsilon_2(\mathbf{k}) & \dots & \sqrt{Z}V_{2M}(\mathbf{k}) \\
\vdots & \dots & \ddots & \dots\\
\hline
 \sqrt{Z}V_{M1}(\mathbf{k}) & \sqrt{Z}V_{M2}(\mathbf{k}) & \dots & Z e_M(\mathbf{k})
\end{array}
\right)
\end{equation}
where $\epsilon_m(\mathbf{k})$ and $V_{mM}(\mathbf{k})$, for $m=1,\dots,M-1$ indicate the dispersions of the $M-1$ bands and their hybridisation\footnote{ In what follows we require that the hybridisation is nonzero almost everywhere, or at least it nullifies in a set of integrable points. The presence of hybridisation is indeed the physical fundamental ground of the present problem.} with the heavy one (labeled $M$), respectively. These parameters implicitly take into account all the renormalization effects other than Z, and tend non-singularly to a finite value even if Z vanishes.

By diagonalising this matrix one indeed gets the final structure of the bands $\Lambda^\nu_{\mathbf{k}}$ at any Z, determining the kinetic energy of the system:
\begin{equation}
\label{eqn:Kinetic_General}
E_{\textrm{kin}}=2 \sum_{\nu\mathbf{k}}\Lambda^\nu_{\mathbf{k}}n_F\left( \Lambda^\nu_{\mathbf{k}} \right),
\end{equation}
This integral is summed over the bands (superscript $\nu$) and over the whole Brillouin zone (subscript $\mathbf{k}$). We argue however that under quite large hypotheses the locus of points $\mathcal{K}_0$ is responsible for the mentioned logarithmic singularity so that we can divide it arbitrarily into:
\begin{equation}
\label{eqn:Kinetic_Sing_Reg}
E_{\textrm{kin}}=E_{\textrm{kin}}^{\textrm{reg}}+E_{\textrm{kin}}^{\textrm{sing}}.
\end{equation}
where $E_{\textrm{kin}}^{\textrm{sing}}$ comes from a slice of k-space including $\mathcal{K}_0$ (i.e. a slice around each Fermi surface sheet of the Z=0 system) and the regular part from the rest of the integral.

\begin{figure}[tb]
\centering
\includegraphics[scale=0.6, keepaspectratio]{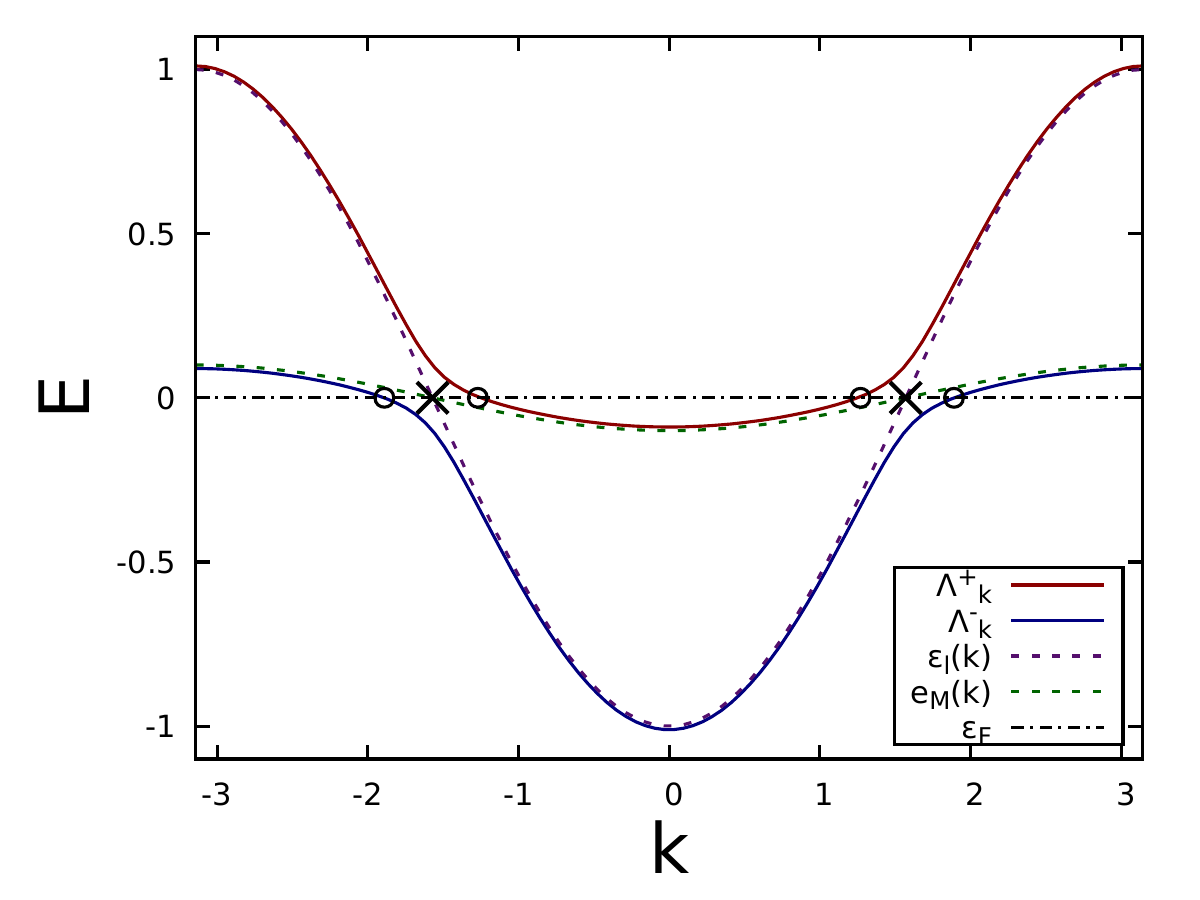}
\caption{\label{fig:hyb_bands} Simplified one-dimensional sketch of a two-band system. Green and purple dashed lines represent two bare, unhybridised bands with dispersion $\epsilon_l(k)$ and $e_M(k)$, respectively. We assume the $M-$th band to be very narrow (renormalised by a small Z), and half-filled. For Z=0 the $e_M(k)$ band flattens on the Fermi level $\epsilon_F$ (black dot-dashed line) while the $\epsilon_l(k)$ band crosses it at cross-marked points. Red and blue solid lines depict the bands in presence of hybridization (also renormalised by $\sqrt{Z}$), crossing the $\epsilon_F$ at the open circles.}
\end{figure}

This is more easily illustrated by a particular two-band example\footnote{We ignore here the points where more than two bands intersect.}, graphically shown in Fig. \ref{fig:hyb_bands}.
In this case the matrix Eq. \ref{eqn:Band_HopMatrix} becomes:
\begin{equation}
\label{eqn:Two_HopMatrix}
\left(
\begin{array}{cc}
\epsilon_l(\mathbf{k}) & \sqrt{Z}V_{lM}(\mathbf{k}) \\
\sqrt{Z}V_{Ml}(\mathbf{k}) & Ze_M(\mathbf{k})
\end{array}
\right).
\end{equation}
If $Z=0$ the flat half-filled band (dot-dashed line) marks the Fermi level and its point cutting $\epsilon_l(\mathbf{k})$ is marked by a black cross, in the uni-dimensional representation of the figure. In three dimensions it will be indeed a (Fermi) surface.

At finite Z one can easily diagonalise the problem to obtain the bands:
\begin{equation}
  \label{eqn:Dressed_Dispersion}
\Lambda^{\pm}_{\mathbf{k}}=\frac{\epsilon_l(\mathbf{k})+Ze_M(\mathbf{k})}{2}\pm R(\mathbf{k})
\end{equation}
where we call:\begin{equation}
R(\mathbf{k})\equiv\sqrt{\left(\frac{\epsilon_l(\mathbf{k})-Ze_M(\mathbf{k})}{2}\right)^2+ZV_{lM}^2(\mathbf{k})}
\end{equation}
represented by the solid lines in Fig.~\ref{fig:hyb_bands}. The main role of $V_{lM}(\mathbf{k})$ is to open a gap between the hybridised bands positioning $\Lambda^+_{\mathbf{k}}$ and $\Lambda^-_{\mathbf{k}}$ respectively always above and below the bare bands. This implies that their crossing points with the Fermi level (circles in Fig.~\ref{fig:hyb_bands}) define a region of $k-$space, which we can call $\Delta \mathcal{K}_0\equiv\left\{\mathbf{k}:\left[ n_F(\Lambda^+_{\mathbf{k}})-n_F(\Lambda^-_{\mathbf{k}})\right]\neq 0\right\}$ encompassing the Z=0 Fermi surface $\mathcal{K}_0$.\\
Now applied to the two bands in Eq. (\ref{eqn:Dressed_Dispersion}), the integral in Eq. (\ref{eqn:Kinetic_General}) is naturally split in two: the part where both bands are occupied or empty at the same time which is regular for $Z\rightarrow 0$ because the cancellation of the radicals $R(\mathbf{k})$, and the complementary one: 
\begin{equation}
\label{eqn:Hyb_K}
E_{\textrm{kin}}^{\textrm{sing}} = \sum_{\mathbf{k}}R(\mathbf{k})\left[ n_F(\Lambda^+_{\mathbf{k}})-n_F(\Lambda^-_{\mathbf{k}})\right]\nonumber\\
=\sum_{\mathbf{k}\in\Delta\mathcal{K}_0}R(\mathbf{k})\to\mathcal{I}(\mathbf{k})\equiv\int_{\mathbf{k}\in\Delta\mathcal{K}_0}\frac{d\mathbf{k}}{(2\pi)^3}R(\mathbf{k}),    
\end{equation}
which we now show is singular.\\
At each point $\mathbf{k}_0\in \mathcal{K}_0$ we can define locally the components of $\mathbf{k}$ parallel ($\mathbf{k}_{\textrm{plane}}(\mathbf{k}_0)$) and perpendicular ($k_{0,\perp}(\mathbf{k}_{0,\textrm{plane}})$) to the surface $\mathcal{K}_0$, and we expand the dispersion in the integrand along the latter direction:
\begin{equation}
\label{eqn:expanded_R}    R(\mathbf{k})=\sqrt{\left(\frac{\nabla_{k_\perp}\epsilon_l- Z\nabla_{k_\perp}e_M}{2}\right)^2(k_\perp -k_{0,\perp})^2+ Z\left(V_0+V_1\nabla_{k_\perp}V_{lM}(k_\perp -k_{0,\perp})\right)^2}    
\end{equation}
where to lighten the notation we define $\nabla_{k_\perp}\epsilon_l\equiv \left. \nabla_{k_\perp}\epsilon_l(\mathbf{k})\right|_{k_\perp = k_{0,\perp}}$, and equivalently for $\nabla_{k_\perp}e_M$ and $\nabla_{k_\perp}V_{lM}$.\\ 
The integral in Eq.~\ref{eqn:Hyb_K} integrates in the radical a polynomial expression in the form $a\Tilde{k}^2+b\Tilde{k}+c$, where $\Tilde{k}\equiv k_\perp - k_{0,\perp}$, $a=\left(\frac{\nabla_{k_\perp}\epsilon_l- Z\nabla_{k_\perp}e_M}{2}\right)^2+ Z\left(V_1\nabla V_{lM}\right)^2$, $b=2 ZV_0V_1\nabla V_{lM}$ and $c= ZV_0^2$. The boundaries of integration ($\chi_{\pm}$, the circles in Fig.~\ref{fig:hyb_bands}) are defined by the nullification condition for the bands in Eq.~\ref{eqn:Dressed_Dispersion}  ($\Lambda^\pm_\mathbf{k}=0$) that shows a crucial property: it is determined by the condition $4\epsilon_l(\mathbf{k})e_M(\mathbf{k})=V_{lM}(\mathbf{k})$ which is \emph{independent} of $ Z$. Thus, $\chi_{\pm}$ remain fixed for $ Z$ going to zero. Vice versa, $a( Z)\to \left(\frac{\nabla_{k_\perp} \epsilon_l}{2}\right)^2$ and $b( Z),c( Z)\to0$ as $ Z\to0$. We now integrate Eq.~\ref{eqn:expanded_R} along the perpendicular direction and then expand for small $Z$. As long as the boundaries of integration are opposite in sign, in the $ Z\to0$ limit we are left with:
\begin{equation}
\label{eqn:log_div}
\mathcal{I}(\mathbf{k})\sim  Z \left[x\sqrt{x^2+1}+\ln\left(x+\sqrt{x^2+1}\right)\right]^{\frac{2a\chi_++b}{\sqrt{4ac-b^2}}}_{\frac{2a\chi_-+b}{\sqrt{4ac-b^2}}}D(\epsilon_l)
\end{equation}
where the integral along the in-plane surface corresponds to the density of states of the bare $\epsilon_l$ dispersion: $D^{Z=0}(E_F)\equiv\int\frac{d\mathbf{k}_{\textrm{plane}}}{(2\pi)^3}(\frac{1}{\nabla_{k_\perp}\epsilon_l})$. In the $ Z\to0$ limit, the first term of Eq.~\ref{eqn:log_div} is finite so that $E_{\textrm{kin}}^{\textrm{sing}}$ in Eq.~\ref{eqn:Kinetic_Sing_Reg} reads as:
\begin{equation}
\label{eqn:diverging_Ekin}
E_{\textrm{kin}}^{\textrm{sing}}=const.+ 2D^{Z=0}(E_F)\bar{V}_0^2 Z \ln ( Z)\,\,\,\,\,\textrm{(Q.E.D.)}.
\end{equation}
where $\bar{V}_0^2$ is the average of $V_0^2$ over $\mathcal{K}_0$.\\

\newpage
\section{Susceptibility measurements on different Cr-doping}
\label{sec:Introduction}

\begin{figure}[h!]
\begin{center}
\includegraphics[width=0.9\textwidth]{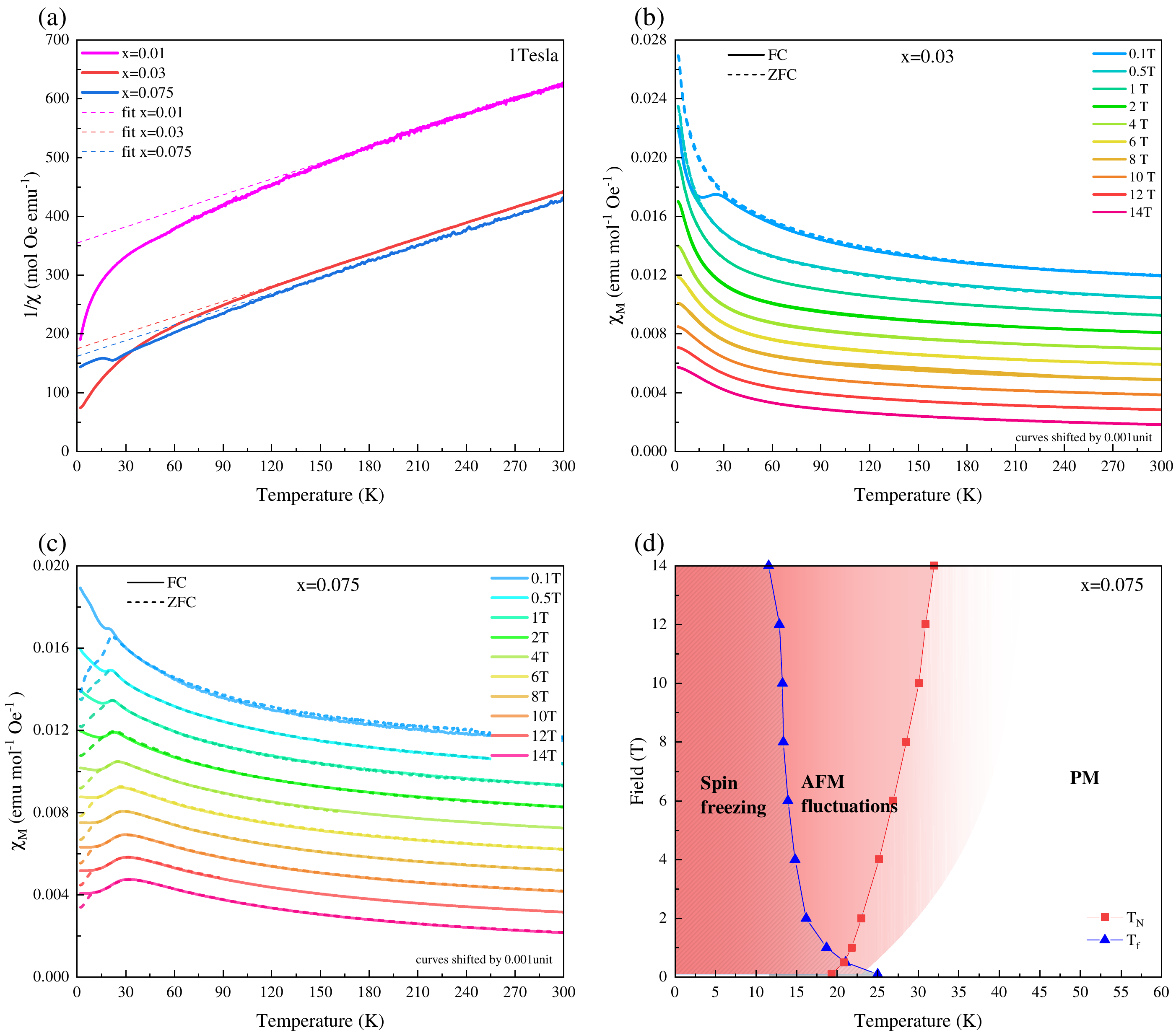}
\caption{(a) Inverse dc magnetic susceptibility of $\mathrm{Cs(Cr_{x}Fe_{1-x})_{2}As_{2}}$ with high-temperature Curie-Weiss fits (dotted lines). (b) and (c) Temperature dependence of the field-cooled (FC) and zero-field-cooled (ZFC) susceptibility for several magnetic-field values, for $x$ = 0.03 and $x$ =0.075, respectively. Curves are shifted for clarity. (d) The resulting phase diagram for $x$ = 0.075, the FWHM of the peak in $\mathrm{d\chi/dT}$ is used to estimate the region of AFM fluctuations. The bifurcation marks the spin freezing temperature.}\label{fig:magnetization supp}
\end{center}
\end{figure}

As shown in Fig.\ref{fig:magnetization supp}(b), a spin-glass-like transition is observed for $x$ = 0.03 around T $\approx$ 40 K as shown by the bifurcation of the zero-field-cooled (ZFC) and field-cooled (FC) curves. It is suppressed for H $\geq$ 2 T where the magnetization becomes fully reversible. For $x$ = 0.075, a competition between spin glass and an AFM phases is observed. The glass transition is shifted to lower temperatures with increasing magnetic field. The ordering temperature of the AFM-like transition (reversible maximum) shifts to higher temperatures. The spin-lattice relaxation in the NMR data reported in Ref.
\onlinecite{CrCs122PRB2023} exhibits a peak around T = 20 K for H = 12 T, supporting the long-range AFM order (also reported for other iron-based superconductors).

\newpage
\section{Thermal expansion}
\label{sec:thermal exp}
\begin{figure}[h!]
\begin{center}
\includegraphics[width=0.85\textwidth]{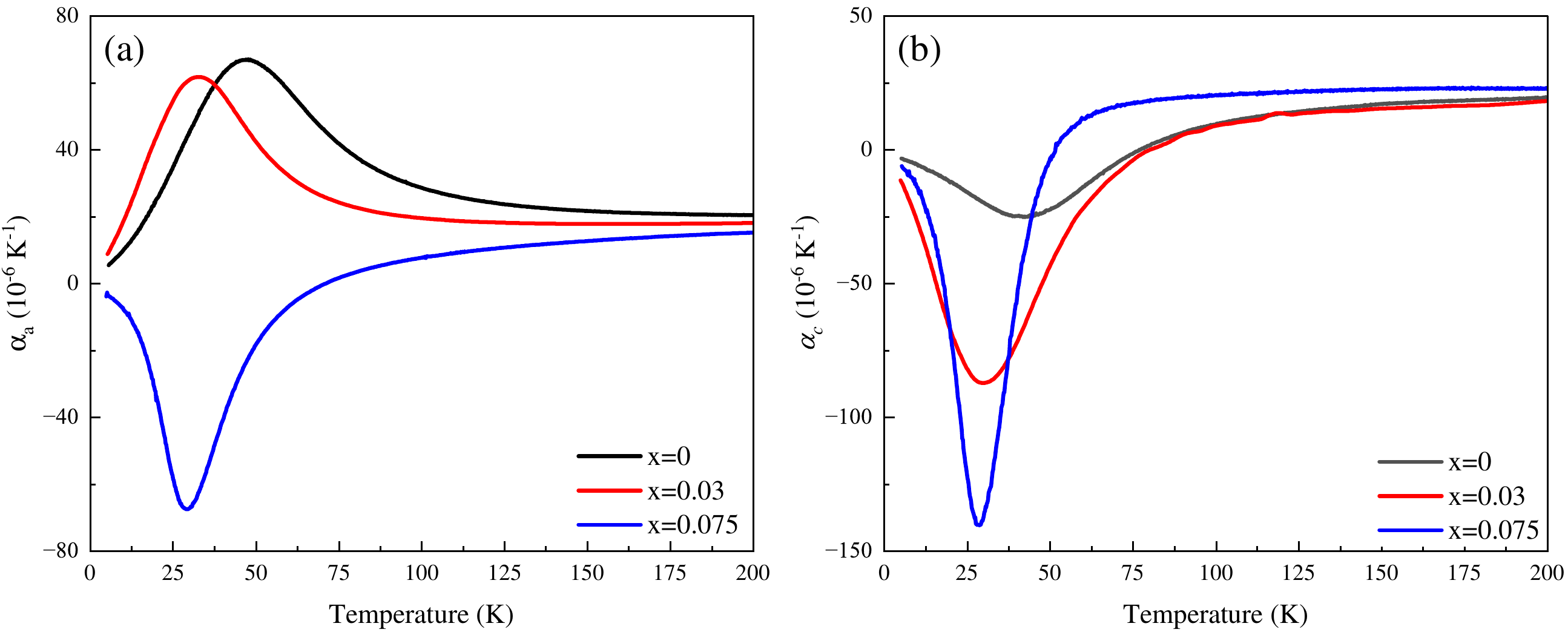}
\caption{(a) and (b) show the respective $a$- and $c$-axis uniaxial thermal expansion.}\label{fig:thermal exp supp}
\end{center}
\end{figure}

\section{Heat Capacity}
\label{sec:heat capacity}
\begin{figure}[h!]
\begin{center}
\includegraphics[width=0.99\textwidth]{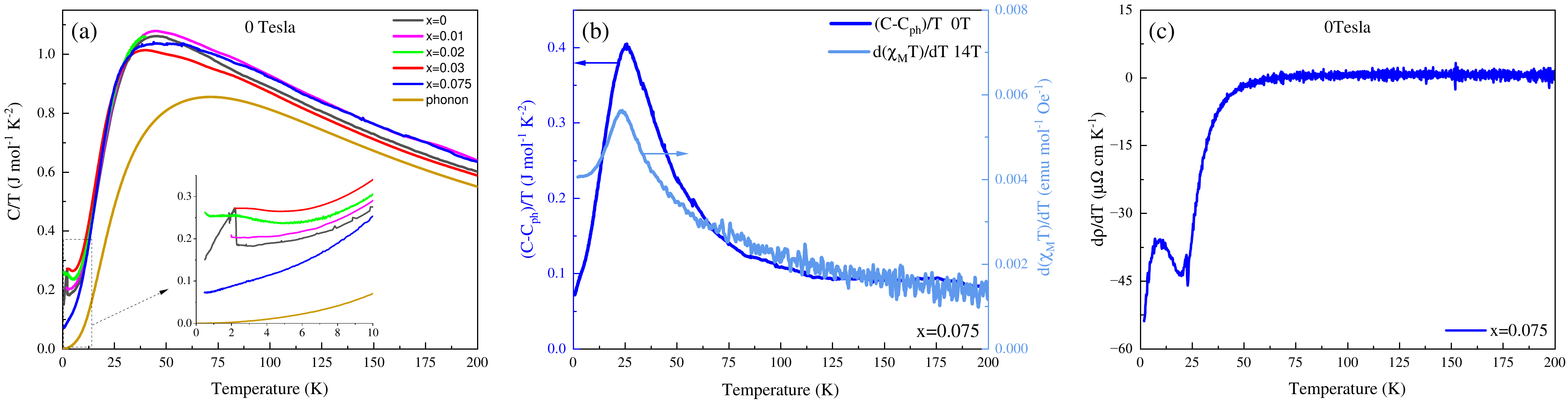}
\caption{ (a) Temperature dependence of the total specific heat $C/T$ for different Cr content together with the phonon contribution obtained from Density-Functional Theory (DFT) calculations.
The inset shows the low-temperature behavior on an enlarged view. (b) Comparison of the electronic specific heat to $d(\chi_MT)/dT$. The latter typically provides a qualitative estimate of the magnetic specific heat of an antiferromagnet in the vicinity of the Néel temperature according to Fisher \cite{magnet-HC-MT}. (c) Temperature derivative of resistivity for $x$ = 0.075 which also exhibits a peak near 25 K.}\label{fig:thermal exp supp}
\end{center}
\end{figure}

\newpage
\section{ Energy dispersive x-ray spectroscopy}
Chemical composition of $\mathrm{Cs(Cr_{x}Fe_{1-x})_{2}As_{2}}$ crystals was determined using energy dispersive x-ray spectroscopy (EDS) in a
COXEM EM-30plus electron microscope equipped with an Oxford Silicon-Drift-Detector (SDD) and AZtecLiveLite-software package. The EDS spectrum
and the elemental mapping of Cr (inset) for $x=0.03$ and $x=0.075$ is shown in Fig.4 (a) and (b).
\begin{figure}[!h]
\centering
\begin{subfigure}{0.48\linewidth}
\centering
\includegraphics[width=1\linewidth, keepaspectratio]{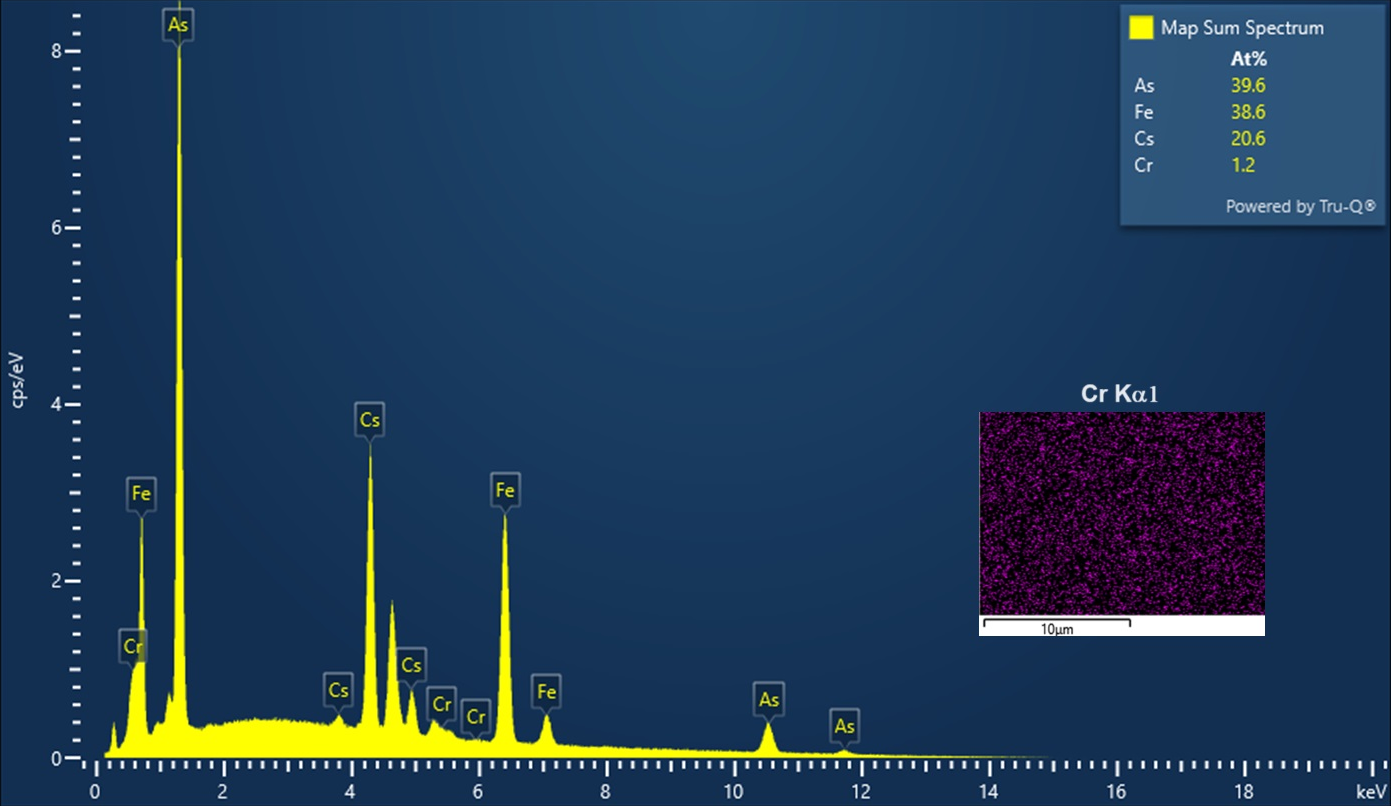}
\caption{}
\label{fig:AAH85_EDX_3doping}
\end{subfigure}
\begin{subfigure}{0.48\linewidth}
\centering
\includegraphics[width=1\linewidth, keepaspectratio]{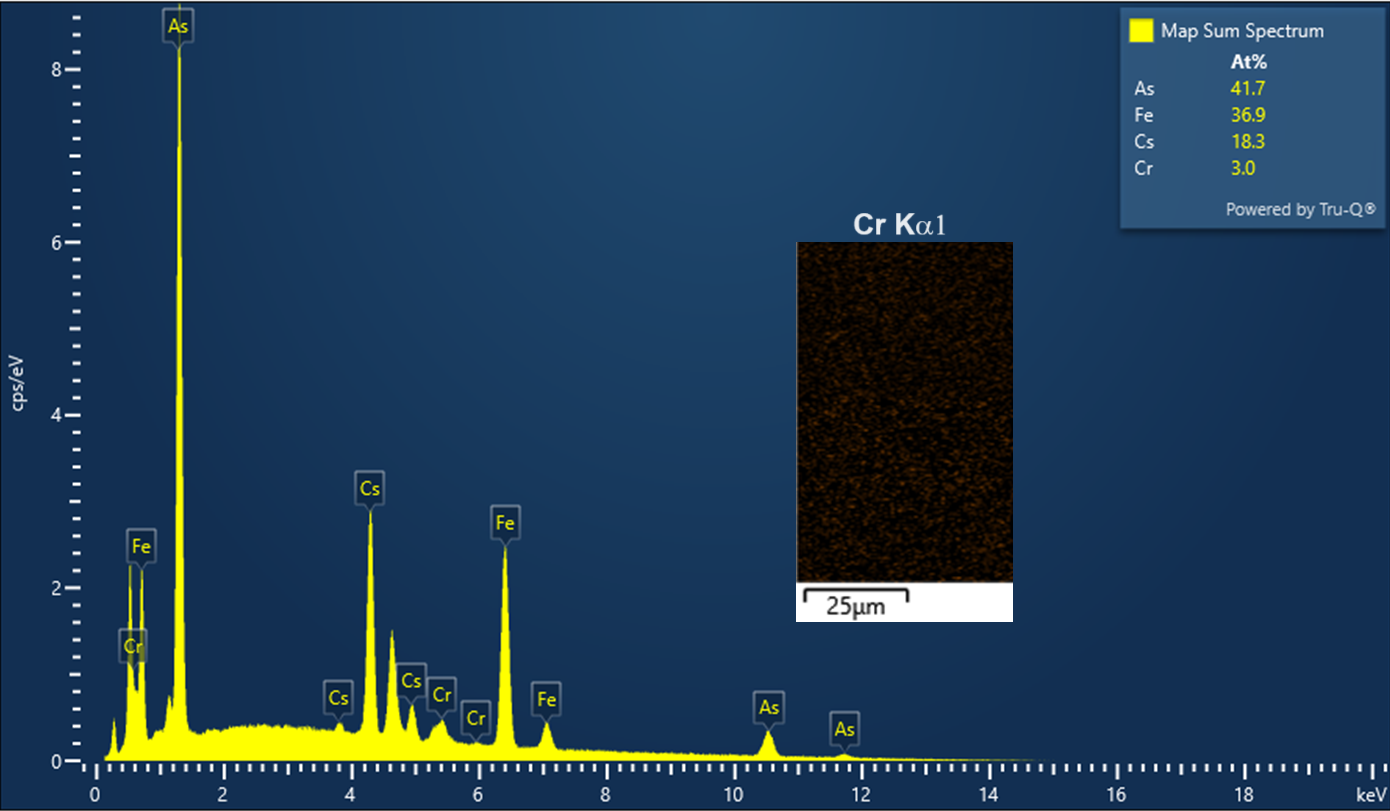}
\caption{}
\label{fig:AAH13_EDX_7doping}
\end{subfigure}
\caption{EDS spectrum for (a) x= 0.03 and (b) x=0.075. The inset shows the uniform distribution of Cr-content in the samples.}
\label{fig:EDX}
\end{figure}


\clearpage
\bibliographystyle{unsrt}
\bibliography{Supp_Refs,suppbib}